\documentclass[11pt,a4paper,twoside,english]{article}
\usepackage{a4}

\usepackage{amssymb,amsmath,mathrsfs}
\usepackage{bm,cancel,units,textcomp,dsfont}
\usepackage{babel} 
\usepackage{color,graphicx}
\usepackage{pictex}
\graphicspath{{abb/}}

\numberwithin{equation}{section}

\begin{document}

\begin{picture}(0,0)
	\put(360,20){ITP-UH-19/07}
\end{picture}

\begin{center}
  {\Large\textbf{Towards the Construction of Local \\[0.5ex]
     Logarithmic Conformal Field Theories \\[1.3ex]}}
\end{center}

\vspace{1cm}

\begin{center}
\textsc{Anne-Ly Do \footnote{email: \texttt{lydo@itp.uni-hannover.de}} and Michael Flohr \footnote{email: \texttt{flohr@itp.uni-hannover.de}}}

  \vspace{1cm}
  {\em Institute for Theoretical Physics, University of Hannover\\
       Appelstra\ss e 2, D-30167 Hannover, Germany} \\ \end{center}

\vspace{1.2cm}

\begin{abstract} 
Although logarithmic conformal field theories (LCFTs) are known not to factorise many previous findings have only been formulated on their chiral halves. Making only mild and rather general assumptions on the structure of an chiral LCFT we deduce statements about its local non-chiral equivalent. Two methods are presented how to construct local representations as subrepresentations of the tensor product of chiral and anti-chiral Jordan cells. Furthermore we explore the assembly of generic non-chiral correlation functions from generic chiral and anti-chiral correlators. The constraint of locality is studied and the generality of our method is discussed. 
\end{abstract}
\newpage
\section{Introduction} 

Since the spadework of Gurarie \cite{Gur93}, an enormous amount of work was done to evolve logarithmic conformal field theories (LCFTs). Their applications to other fields in physics, for example to the theory of percolation and critical disordered systems have been explored. Many structural aspects have been studied in detail. In particular the powerfull techniques from non-logarithmic CFT have been ported to LCFT. The decoupling of the conformal symmetry algebra in two independent sectors of opposite chirality gives reason to a especially elegant and beneficial technique known from the non-logarithmic case: Many statements about CFTs can be derived considering only one sector, either the chiral or the anti-chiral. This means that objects are considered which transform effectively in a left- \textit{or} right-handed representation. In case of such a constriction on ``half'' objects it is common to speak about chiral theories. Results of one sector can be transferred to the other one by mirroring the chirality. Subsequently both halves have to be glued together to form the full, non-chiral theory.   

Among the findings which were up to now stated for chiral LCFTs only, the following topics are of central interest: Correlation functions are calculated for instance in \cite{Flo00, Flo01, GheKa97, RAK96, RA05, RA06, FloK05, FloK03, MRS02, Mathieu07}, fusion rules are investigated, among others, in \cite{GabK96, Gab01, RasPear07, EF06, FloKn07} and some studies on null vectors can be found in \cite{Flo97, EbF05}. Enlarging their scope to non-chiral theories and implementing the constraints of locality is a task of elementary importance, as only local theories have physical interpretation. Unfortunately, the construction of local theories turned out to be non-trivial: Unlike conventional, non-logarithmic CFTs, LCFTs were found to be non-factorisable. Insofar speaking about chiral halves in the context of LCFTs might be delusive. As we will see, assembling a local theory out of a chiral LCFT and its anti-chiral counterpart is evocative of screwing them into each other rather than of combining two halves. Anyhow, we will stick with the familiar naming convention and refer in abuse of language to the chiral theories as chiral halves.     

So far, only few attempts have been made to close the gap between the well-known chiral and the almost unknown non-chiral LCFTs. Gaberdiel and Kausch succeeded in constructing a non-chiral local theory at $c=-2$ by solving the conformal bootstrap \cite{GabK99}. This construction was interpreted in terms of symplectic fermions \cite{Kau00} and enabled a detailed comparison of the two-dimensional Abelian sandpile model with the local triplet theory at $c=-2$ \cite{JPR06}. The logarithmic triplet theory with boundary was studied in \cite{GabR06} for $c_{1,2}=-2$ and the generalisation to rational $c_{1,p}$ models was achieved in \cite{GabR07}.
Stating remarkable structural similarities between the local triplet theory and supergroup WZNW models, Schomerus et al. suggested that consulting the better understood local WZNW models might promote the construction of generic local LCFTs. Local logarithmic bulk correlation functions for the $GL(1|1)$ WZNW model have been computed in \cite{SaSch05}, non-chiral indecomposable representations on which the zero-mode of the energy-momentum tensor is not digonalizable were investigated by means of the WZNW model on the supergroup $PSU(1,1|2)$ \cite{GQuSch06}. A rather general discussion on conclusions for local LCFTs from the supergroup WZNW point of view can be found in \cite{QuSch07}.

In this paper, we will choose an approach which rests solely upon the analysis of the conformal symmetry and aims for a high degree of generality. Beside some rather general assumptions on the structure of an chiral LCFT the deduced statements about its local non-chiral equivalent can be called generic.

The paper is organized as follows: In section 2 we fix the naming convention. Section \ref{content_sec} presents two methods how to construct the space of states of a local LCFT. Both methods predict a non-chiral local theory to possess the same rank as the halves it is composed of. In other respects the results of both methods turn out to be mutually excluding. The consideration of a specific model suggests one of the proposed methods to be more intuitive. From section \ref{cor_sec} we turn our attention to LCFT correlation functions. After a brief recapitulation about generic chiral correlation functions, we explore the possibility to compose generic non-chiral correlators out of generic chiral ones. Beside the constraint of locality, invariance under the global conformal group, duality and monodromy invariance have to be implemented. We propose a construction method, discuss the generality of our solution and check its consistency with previous findings. More details on this work can be found in \cite{Ly07}.   
  
\section{Definitions and Preliminaries}

Logarithmic conformal field theories feature indecomposable but reducible representations of the chiral symmetry algebra. In this paper we will consider only the case where such indecomposable representations occur with respect to the Virasoro zero modes. Through appropriate choice of the basis, the generators $L_{0}$ and $\overline{L_{0}}$ can be transformed into Jordan normal form. A rank $r$ Jordan cell is spanned by $r$ fields $\left\{\Psi_{(h,r-1)},\ldots,\Psi_{(h,1)},\Psi_{(h,0)}\right\}$. The action of the zero mode $L_{0}$ of the Virasoro algebra is then given by 
	\[
	L_{0}\Psi_{(h,k)}(0)|0\rangle=h\:\Psi_{(h,k)}(0)|0\rangle+(1-\delta_{k,0})\Psi_{(h,k-1)}(0)|0\rangle,
\]
where $h$ is the conformal weight and $|0\rangle$ as usual denotes the $SL(2,\mathbb{C})$ invariant vacuum. The parameter $k$ grades the fields within the Jordan cell and will be referred to as Jordan level. In the scope of this paper, only reducible representations are regarded whose irreducible subrepresentations  $\Psi_{(h_{i},0)}(z_{i})$ accord to proper primaries, i.\,e.\ to fields whose OPE among each other does never produce logarithmic fields on the right hand side. Furthermore, we assume the logarithmic partner fields of the proper primary to be quasi-primary, i.e. $L_{n}\Psi_{(h_{i},k_{i})}=0\quad\forall n\geq1$. 
Of course, our assumptions limit the generality with which our results are
valid. This is particularly true for our assumption on the quasi-primarity of
the logarithmic partner fields. 
We will discuss these limitations and how they may be overcome in the
conclusions. Thus, our paper should be understood as a first step towards
a full treatment of the question of local logarithmic conformal field theories.

If an LCFT accomodates more than one indecomposable representation one can consider every Jordan cell to be of rank $r(h)=r$, with $r$ being the rank of the largest Jordan cell. Hypothetically emerging smaller Jordan cells can be padded with fields that are to be set zero afterwards.  

The following decomposition of the non-diagonal action of the Virasoro modes on LCFT $n$-point correlators will turn out to be advantageous:
\begin{equation}\label{Viraction_gl}
L_{q}\left\langle\ldots\right\rangle=\left[O_{q}+\sum^{n}_{i=1} z^{q}_{i}(q+1)\delta_{h_{i}}\right]\left\langle\ldots\right\rangle,
\end{equation}
where $O_q$ abbreviates the diagonal part of the action as known from ordinary non-logarithmic CFT
	\[O_{q}\left\langle\ldots \right\rangle=\sum^{n}_{i=1} z^{q}_{i}\left[z_{i}\partial_{i}+(q+1)h_{i}\right]\left\langle\ldots \right\rangle.
\]
The off-diagonal, nilpotent part is generated by operators $\delta_{h_{i}}$. They act on a logarithmic field by reducing its Jordan level by one, on a primary by annihilating the field:	
\[
\delta_{h_{j}}\Psi_{(h_{i},k_{i})}(z_{i})=\delta_{ij}(1-\delta_{0,k_{i}})\Psi_{(h_{i},k_{i-1})}(z_{i}).
\] 

Below we will augment the introduced glossary with entities marked with a bar. These can be obtained from those without a bar by complex conjugation of all variables $z$ and providing all parameters $h$, $k$ and $g$ with a superscript line. For $h$, $k$ and $g$ this line is not related to complex conjugation, but only indicates that parameters of the anti-chiral theory are denoted. 

To distiguish quantities of a full local theory from those living on its right- or lefthanded half we will apply the pair of concepts ``non-chiral'' - ``chiral''. If used in this sense the latter shall cover anti-chiral quantities, too.

\subsection{Locality constraints}
\label{sec:Locality-constraints}

The fundamental postulation on a non-chiral theory is locality of the fields or, in terms of correlation functions, singlevaluedness of the amplitudes. In \cite{GabK99} it was shown that this imposes the following constraints: 
\begin{equation}\label{LC_gl} 
 h_{i}-\bar{h_{i}}\in\mathbb Z,\qquad \wedge \qquad S_{i}\Psi_{i}=0, 
\end{equation}
where $\Psi_{i}$ abbreviates $\Psi_{(h_{i},\bar{h}_{i},\mathbf{k_{i}})}$ and $S_{i}$ shortens $\delta_{h_{i}}-\delta_{\bar{h}_{i}}$. Here we introduced $\mathbf{k_{i}}$ anticipatorily as non-chiral Jordan level. As the notion Jordan level rests upon the Jordan block structure of the zero modes of the chiral symmetry algebra, we need a redefinition in terms of non-chiral representations, which will be given in section \ref{content_sec}.

\section{Non-chiral local representations}\label{content_sec}

Non-chiral irreducible representations can be obtained as diagonal tensor product of irreducible chiral representations \cite{DiFr97}. This course of action fails in case of non-chiral indecomposable representations. A chiral indecomposable representation $\mathcal{R}_{h}$ of Jordan rank $r+1$ is generated by states $\left|h,k_{i}\right\rangle$, $k_{i}$ running from $r$ to null. The various tensor products included in $\mathcal{R}_{h}\otimes\mathcal{R}_{\bar{h}}$ can be endowed with a gradation: Starting with the tensor product of those states with highest Jordan level, all other possible tensor products can be obtained by repeated application of $\delta_{h}$ and $\delta_{\bar{h}}$. The resulting structure is summarised in figure 1: 
\vspace*{10mm}
\begin{center}
\setlength{\unitlength}{.92pt}
\begin{picture}(110,170)
\label{sc_fig}
  \put(200,163){$\mathbf{r}$}
  \put(0,163){$r\otimes\bar{r}$} 
  \put(15,156){\vector(1,-1){20}}
  \put(13,156){\vector(-1,-1){20}} 
  \put(190,123){$\mathbf{r-1}$} 
  \put(-50,123){$(r\negthickspace-\negthickspace1)\otimes\overline{r}$} 
  \put(27,123){$r\otimes(\overline{r\negthickspace-\negthickspace1})$}        
  \put(-15,117){\vector(1,-1){20}}
  \put(-18,117){\vector(-1,-1){20}}  
  \put(41,117){\vector(-1,-1){20}}
  \put(43,117){\vector(1,-1){20}}
  \put(190,85){$\mathbf{r-2}$}
  \put(-83,85){$(r\negthickspace-\negthickspace2)\otimes\overline{r}$}              \put(-23,85){$(r\negthickspace-\negthickspace1) \otimes(\overline{r\negthickspace-\negthickspace1})$} 
  \put(58,85){$r\otimes(\overline{r\negthickspace-\negthickspace2})$} 
  \put(202,55){\vdots}
  \put(11,55){\vdots}
  \put(200,30){$\mathbf{0}$}
  \put(-103,30){$0\otimes\overline{r}$} 
  \put(-53,30){$1\otimes(\overline{r\negthickspace-\negthickspace1})$} 
  \put(101,30){$r\otimes\overline{0}$} 
  \put(30,30){$(r\negthickspace-\negthickspace1)\otimes\overline{1}$}   
  \put(5,30){$\cdots$}
  \put(-88,25){\vector(1,-1){20}}
  \put(-40,25){\vector(-1,-1){20}} 
  \put(-38,25){\vector(1,-1){20}}
  \put(65,25){\vector(-1,-1){20}} 
  \put(67,25){\vector(1,-1){20}}
  \put(111,25){\vector(-1,-1){20}}
  \put(197,-5){$\mathbf{-1}$}
  \put(-79,-5){$0\otimes(\overline{r\negthickspace-\negthickspace1})$} 
  \put(53,-5){$(r\negthickspace-\negthickspace1)\otimes\overline{0}$}   
  \put(5,-5){$\cdots$}
  \put(-62,-19){$\ddots$}
  \put(78,-12){.}
  \put(74,-16){.}
  \put(70,-19){.}
  \put(190,185){\textbf{level}}
  \put(202,-26){\vdots}
  \end{picture}
  \vspace*{7mm}
  \end{center} 
\begin{center}
Figure 1: \textit{Gradation of $\mathcal{R}_{h}\otimes\mathcal{R}_{\bar{h}}$}\phantom{XXXXXXXX}
\end{center}
     
Here the following abbreviations were introduced: An element of the product space $\mathcal{R}_{h}\otimes\mathcal{R}_{\bar{h}}$ is uniquely denoted by its chiral and anti-chiral Jordan level  
	\[k\otimes\bar{k'}:=|h,k\rangle\otimes|\bar{h},\bar{k'}\rangle.
\]
Let $\swarrow$ indicate that the state the arrow points at is an image of the state the arrow starts at under the action of $\delta_{h}$. The symbol $\searrow$ does the same for the action of $\delta_{\bar{h}}$. Level $n=r-m$ is subsequently assigned to states that are obtained from $r\otimes\bar{r}$ by $m$-fold descending, i.\,e.\ by $m$-fold application of delta operators. Starting with one state on level $r$, the number of states with level $n$ increases gradually with decreasing $n\geq0$. Level zero possesses the highest number of states, namely $r+1$. For level $n\leq0$ those states located on the fringes of the above graphic are deleted either by $\delta_{h}$ or by $\delta_{\bar{h}}$, hence the number of states on level $n$ decreases with decreasing $n\leq0$:   
	\[\mbox{number of states in $\mathcal{R}_{h}\otimes\mathcal{R}_{\bar{h}}$ with level n}\ \, = \ \, r+1-\left|n\right|, \qquad  \left|n\right|\leq r.
\]

Due to the field-state isomorphism $|h,k\rangle:=\lim_{z\rightarrow0}\Psi_{(h,k)}(z)|0\rangle$ equation (\ref{LC_gl}) enforces that the action of the operator $S=\delta_{h}-\delta_{\bar{h}}$ has to vanish on all non-chiral states belonging to local theories. Figure 1 shows that generally the image of a tensor state under the action of $\delta_{h}$  differs from its image under $\delta_{\bar{h}}$. As a consequence states of the local non-chiral theory span only a subspace of the space generated by the diagonal tensor product of the chiral and anti-chiral Jordan cells \cite{GabK99}. This local subspace $\mathcal{L}_{h\bar{h}}$ can be constructed in two ways. Firstly, we will revisit the ``quotient space construction'' which Gaberdiel and Kausch introduced for the $c=-2$ local triplet theory \cite{GabK99} and enlarge the scope of their findings to theories whose chiral halves are of arbitrary rank. We would like to point out that a generalisation was achieved earlier but under a different viewpoint: Starting off at a single boundary condition Gaberdiel and Runkel constructed the compatible local space of bulk states \cite{GabR07}. Their course of action holds for rational CFTs, logarithmic and non-logarithmic, and was shown to reproduce the known local bulk theory at $c=-2$ if applied to the $c_{1,2}$ triplet model. Our approach towards an enhanced quotient space construction differs from \cite{GabR07} inasmuch it does not use any information about boundaries. 

Secondly, we will introduce the ``kernel construction'' and compare the results of both methods.  
\enlargethispage{\baselineskip}
\subsection{Quotient space construction}

Locality requires that the image of a local state under the action of $\delta_{h}$ has to equal the image of the same state under the action of $\delta_{\bar{h}}$ (\ref{LC_gl}). The ``quotient space construction'' presented here approaches the problem by identifying both images modulo elements of a subspace $\mathcal{N}_{h\bar{h}}$:  
\begin{equation}	\mathcal{L}_{h\bar{h}\;QSC}\equiv\left(\mathcal{R}_{h}\otimes\mathcal{R}_{\bar{h}}\right)/\mathcal{N}_{h\bar{h}}.
\end{equation}
The chiral version of the $c=-2$ triplet theory possesses two indecomposable representations of rank two. For these cases, Gaberdiel and Kausch identified the subrepresentation spanned by $S$ acting on the tensor product of states with Jordan level one and the descendants of this state with respect to $\delta_{h}$ and $\delta_{\bar{h}}$ to be a minimal choice for $\mathcal{N}_{h\bar{h}}$ \cite{GabK99}:
	\[\mathcal{N}_{h\bar{h}}=\left\{(\delta_{h})^{p}(\delta_{\bar{h}})^{q}S(1\otimes\bar{1})\right\}, \quad h=\bar{h}=0,1.
\]
This result can be generalised to local theories whose chiral halves exhibit Jordan cells of arbitrary rank $r+1$. For that purpose we repeat the procedure used to gain the gradation of $\mathcal{R}_{h}\otimes\mathcal{R}_{\bar{h}}$ with a different starting point: $S$ acting on the tensor product of those states with highest Jordan level. The structure of the resulting subrepresentation $S_{h\bar{h}}$ is depicted on the right hand side of the chart given below. 

\vspace*{2mm}
\begin{center}
\setlength{\unitlength}{.92pt}
\begin{picture}(210,160)
  \put(11,158){$\bullet$} 
  \put(15,153){\vector(1,-1){20}}
  \put(13,153){\vector(-1,-1){20}}  
  \put(-17,122){$\bullet$} 
  \put(38,122){$\bullet$}        
  \put(-15,118){\vector(1,-1){20}}
  \put(-18,118){\vector(-1,-1){20}}  
  \put(43,118){\vector(-1,-1){20}}
  \put(45,118){\vector(1,-1){20}}
  \put(-46,87){$\bullet$}       
  \put(11,87){$\bullet$} 
  \put(68,87){$\bullet$} 
  \put(11,57){\vdots}
  \put(-86,35){$\bullet$} 
  \put(99,35){$\bullet$}   
  \put(-30,35){$\bullet$} 
  \put(45,35){$\bullet$}    
  \put(5,35){$\cdots$}
  \put(-78,30){\vector(1,-1){20}}
  \put(99,30){\vector(-1,-1){20}}
  \put(-32,30){\vector(-1,-1){20}}
  \put(-23,30){\vector(1,-1){20}}  
  \put(53,30){\vector(1,-1){20}}
  \put(44,30){\vector(-1,-1){20}}
  \put(-58,0){$\bullet$} 
  \put(73,0){$\bullet$}   
  \put(5,0){$\cdots$}
  \put(-50,-5){\vector(1,-1){20}}
  \put(-2,-5){\vector(-1,-1){20}}
  \put(24,-5){\vector(1,-1){20}}
  \put(71,-5){\vector(-1,-1){20}}
  \put(-29,-34){$\bullet$} 
  \put(44,-34){$\bullet$}   
  \put(5,-34){$\cdots$}
  \put(-24,-40){.}
  \put(-20,-44){.}
  \put(-16,-48){.}
  \put(40,-40){.}
  \put(36,-44){.}
  \put(32,-48){.}  
  \put(75,156){\vector(4,-1){80}}
  \put(120,156){$S$}
  \put(211,122){$\circ$}
  \put(215,118){\vector(1,-1){20}}
  \put(213,118){\vector(-1,-1){20}} 
  \put(183,87){$\circ$}       
  \put(240,87){$\circ$}
  \put(212,57){$\vdots$} 
  \put(154,35){$\circ$}   
  \put(208,35){$\cdots$}
  \put(269,35){$\circ$} 
  \put(156,30){\vector(-1,-1){20}}
  \put(159,30){\vector(1,-1){20}}
  \put(204,30){\vector(-1,-1){20}}  
  \put(226,30){\vector(1,-1){20}}
  \put(270,30){\vector(-1,-1){20}}
  \put(273,30){\vector(1,-1){20}}
  \put(132,0){$\circ$}
  \put(179,0){$\circ$} 
  \put(208,0){$\cdots$}
  \put(246,0){$\circ$}
  \put(292,0){$\circ$}
  \put(135,-5){\vector(1,-1){20}}
  \put(180,-5){\vector(-1,-1){20}}
  \put(183,-5){\vector(1,-1){20}}
  \put(248,-5){\vector(-1,-1){20}}
  \put(251,-5){\vector(1,-1){20}}
  \put(295,-5){\vector(-1,-1){20}}
  \put(155,-34){$\circ$}
  \put(208,-34){$\cdots$}
  \put(270,-34){$\circ$}
  \put(160,-40){.}
  \put(164,-44){.}
  \put(170,-48){.}
  \put(266,-40){.}
  \put(262,-44){.}
  \put(258,-48){.}    
  \put(-10,-75){$\mathcal{R}_{h}\otimes\mathcal{R}_{\bar{h}}$}
  \put(205,-75){$\mathcal{S}_{h\bar{h}}$}
  \setlength{\unitlength}{1pt}
\end{picture}
 \end{center} 
\vspace*{18mm}
\begin{center}
Figure 2: \textit{$\mathcal{R}_{h}\otimes\mathcal{R}_{\bar{h}}$ versus $\mathcal{S}_{h\bar{h}}$}
\end{center}
\vspace{\baselineskip}

On the left hand side the gradation of $\mathcal{R}_{h}\otimes\mathcal{R}_{\bar{h}}$ is recapitulated: Every $\bullet$ marks an element of the product space. As per construction the formation on the right hand side is generated by one state $S(r\otimes\bar{r})=(r\negthickspace-\negthickspace1)\otimes\overline{r}-r\otimes(\overline{r\negthickspace-\negthickspace1})$ on level $r-1$. This state and every state that emanates from it by application of delta operators are pictured as $\circ$. Counting the states $\in\mathcal{S}_{h\bar{h}}$ with level n yields   
 	\[\mbox{number of states in $\mathcal{S}_{h\bar{h}}$ with level n} \ \,=\ \, r+1-\left|n+1\right|, \qquad  \left|n\right|\leq r-1.
\]
Level $n=-1$ accomodates the maximum number of states. Level $n\leq-1$ states that are located on the fringes vanish either under the action of $\delta_{h}$ or of $\delta_{\bar{h}}$. 

Comparing the state content of both representations one finds: For $\left|n\right|\leq r-1$, the difference of two adjacent level $n$ states in $\mathcal{R}_{h}\otimes\mathcal{R}_{\bar{h}}$ is element of $S_{h\bar{h}}$. Additionally, for level $n\leq-1$ the skirting states of both representations are pairwise identical:  

\vspace*{5mm}
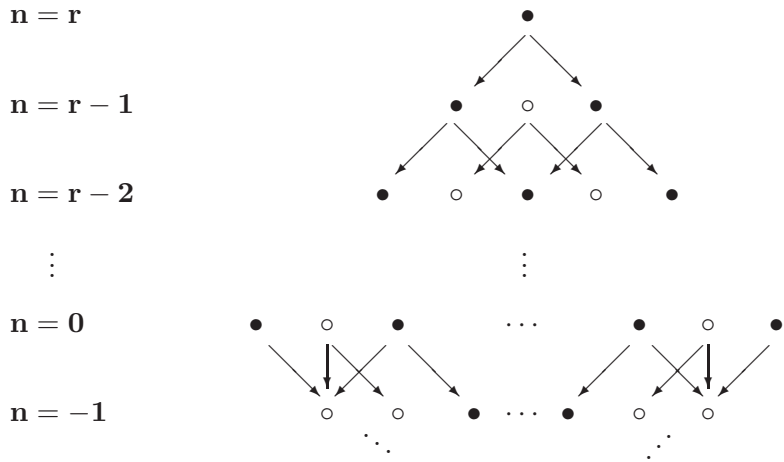
\begin{center}
\setlength{\unitlength}{.95pt}
\begin{picture}(10,160)
  \put(-190,158){$\mathbf{n=r}$}
  \put(11,158){$\bullet$} 
  \put(15,153){\vector(1,-1){20}}
  \put(13,153){\vector(-1,-1){20}}  
  \put(-190,122){$\mathbf{n=r-1}$} 
  \put(-17,122){$\bullet$} 
  \put(11,122){$\circ$}
  \put(38,122){$\bullet$}       
  \put(-190,87){$\mathbf{n=r-2}$} 
  \put(-15,118){\vector(1,-1){20}}
  \put(-18,118){\vector(-1,-1){20}}  
  \put(15,118){\vector(1,-1){20}}
  \put(13,118){\vector(-1,-1){20}}
  \put(43,118){\vector(-1,-1){20}}
  \put(45,118){\vector(1,-1){20}}
  \put(-46,87){$\bullet$}
  \put(-17,87){$\circ$}       
  \put(11,87){$\bullet$} 
  \put(38,87){$\circ$}
  \put(68,87){$\bullet$} 
  \put(11,57){\vdots}
  \put(-96,35){$\bullet$} 
  \put(-68,35){$\circ$}
  \put(-40,35){$\bullet$} 
  \put(5,35){$\cdots$}
  \put(55,35){$\bullet$} 
  \put(82,35){$\circ$}
  \put(109,35){$\bullet$}   
  \put(-175,57){\vdots}
  \put(-190,35){$\mathbf{n=0}$}
  \put(-88,30){\vector(1,-1){20}}
  \put(-63,30){\vector(1,-1){20}}
  \put(-65,30){\vector(0,-1){17}}
  \put(-42,30){\vector(-1,-1){20}}
  \put(-33,30){\vector(1,-1){20}} 
  \put(54,30){\vector(-1,-1){20}} 
  \put(63,30){\vector(1,-1){20}}
  \put(83,30){\vector(-1,-1){20}}
  \put(85,30){\vector(0,-1){17}}
  \put(109,30){\vector(-1,-1){20}}
  \put(-190,0){$\mathbf{n=-1}$}
  \put(-68,0){$\circ$}
  \put(-40,0){$\circ$}
  \put(-10,0){$\bullet$}
  \put(5,0){$\cdots$}
  \put(27,0){$\bullet$}
  \put(55,0){$\circ$} 
  \put(82,0){$\circ$}   
  \put(-52,-14){$\ddots$}
  \put(68,-7){.}
  \put(64,-11){.}
  \put(60,-15){.}
\setlength{\unitlength}{1pt}
 \end{picture}
 \end{center} 
\vspace*{3mm}
\begin{center}
Figure 3: \textit{Comparison of $\mathcal{R}_{h}\otimes\mathcal{R}_{\bar{h}}$ and $\mathcal{S}_{h\bar{h}}$}
\end{center}
\vspace{.5\baselineskip}

Hence dividing the subspace $\mathcal{S}_{h\bar{h}}$ out of $\mathcal{R}_{h}\otimes\mathcal{R}_{\bar{h}}$ accords with identification of all those states $\in \mathcal{R}_{h}\otimes\mathcal{R}_{\bar{h}}$ that exhibit identical level:
	\[(\delta_{h})^{p_{i}}(\delta_{\bar{h}})^{q_{i}}\:r\otimes \bar{r}\sim(\delta_{h})^{p_{j}}(\delta_{\bar{h}})^{q_{j}}\:r\otimes \bar{r}\qquad \forall i,j|p_{i}+q_{i}=p_{j}+q_{j}.
\]
The action of $\delta_{h}$ on these equivalence classes equals that of $\delta_{\bar{h}}$. Thus indeed with $S_{h\bar{h}}$ we constructed a minimal choice for $\mathcal{N}_{h\bar{h}}$. 

The equivalence classes in $\left(\mathcal{R}_{h}\otimes\mathcal{R}_{\bar{h}}\right)/\mathcal{S}_{h\bar{h}}$ are parameterised and arranged in order by their level. A non-chiral level $n$ state shall be defined as a representative of the equivalence class with level $n$. As standard representative we choose the symmetric sum over all elements:
\begin{equation}\label{standrep_gl}
	\left|h,\bar{h},\mathbf{n}\right\rangle:=\sum\text{level $n$ states}.
\end{equation}
To avoid confusion with the Jordan level naming of a chiral state, we will use bold numbers to denote states of the non-chiral theory by their level. Since the equivalence classes with level $n\leq-1$ include representatives $\in\mathcal{S}_{h\bar{h}}$, they are entirely removed from $\left(\mathcal{R}_{h}\otimes\mathcal{R}_{\bar{h}}\right)/\mathcal{S}_{h\bar{h}}$, i.\,e.\ states $\in\mathcal{R}_{h\bar{h}}$ at level zero are annihilated by both delta operators.

\subsection{Kernel construction}
The ``kernel construction'' defines $\mathcal{L}_{h\bar{h}\;KC}\subset\mathcal{R}_{h}\otimes\mathcal{R}_{\bar{h}}$ as the kernel of $S$. The key idea for determining the kernel of $S$ is to use telescoping series. If we sum up all states $\in \mathcal{R}_{h}\otimes\mathcal{R}_{\bar{h}}$ with same level and act with $S$ on it, every term except the first and the last cancels with either the preceding or suceeding term. 
	\[S\left(\sum\text{level $n$ states}\right)=\begin{cases}\delta_{h}\left(n\otimes\bar{r}\right)-\delta_{\bar{h}}\left(r\otimes\bar{n}\right)\qquad\qquad\qquad\text{for}\;n\geq1\\
\delta_{h}\left(0\otimes(r-n)\right)-\delta_{\bar{h}}\left((r-n)\otimes\bar{0}\right)\quad\text{for}\;n\leq0.
	\end{cases}
\]

As for level $\leq0$ the surviving terms vanish, we can conclude:    
	\[\mathcal{L}_{h\bar{h}\;KC}=\left\{\sum\text{level $n$ states}\:|\:n\leq0\right\}.
\]

\subsection{Discussion and summary}\label{loccondiss_sec}

In the preceding sections we presented two methods how to construct local indecomposable representations - the quotient space construction (QSC) and the kernel construction (KC). According to both methods $\mathcal{L}_{h\bar{h}}$ possesses the same rank as the chiral halves it is composed of:
	\[\text{rank}(\mathcal{L}_{h\bar{h}})=\text{rank}(\mathcal{R}_{h})=\text{rank}(\mathcal{R}_{\bar{h}}). 
\]
The state content of a local representation depends on the method used to construct it:
	\[
	\begin{split}
	&\mathcal{L}_{h\bar{h}\;QSC}=\left\{\sum\text{level $n$ states}\:|\:n\geq0\right\}\\
&\mathcal{L}_{h\bar{h}\;KC}\;\,=\left\{\sum\text{level $n$ states}\:|\:n\leq0\right\}
\end{split}
\]
Above, specifications were made such that for QSC the term non-chiral Jordan level is in perfect accordance to the chiral Jordan level. For KC it is of avail to slightly adapt the definition of the non-chiral Jordan level: Redefining level $n$ to be level $n+r$
	\[n\mapsto n'=n+r
\]
we achieve the familiar situation of non-negative integer values for the non-chiral Jordan level. Furthermore the redefinition guarantees that level zero states are annihilated by $\delta_{h}$ and $\delta_{\bar{h}}$: 
\[\delta_{h}\left|h,\bar{h},\mathbf{n'}\right\rangle_{KC}=\delta_{\bar{h}}\left|h,\bar{h},\mathbf{n'}\right\rangle_{KC}=0\quad \text{for}\; \mathbf{n'}=\mathbf{0}.
\]
 
Computable predictions in QSC may depend on the choice of the considered representative. The vanishing Shapovalov form of $|\mathbf{0}\rangle$ for example can be shown by means of the representatives $0\otimes\bar{r}$ and $r\otimes\bar{0}$. However, rank $r+1$ theories with $r+1>1$ and $r$ an even integer possess a level zero representative $\tfrac{r}{2}\otimes\bar{\tfrac{r}{2}}$. According to equation \eqref{Anfbed_gl} the corresponding Shapovalov form is   
	\begin{equation}
	\Bigl(\left\langle \vphantom{\bar{\tfrac{r}{2}}}\tfrac{r}{2}\right|\otimes\left\langle \bar{\tfrac{r}{2}}\right|\Bigr)\;\Bigl(\left| \vphantom{\bar{\tfrac{r}{2}}}\tfrac{r}{2}\right\rangle\otimes\left|\bar{\tfrac{r}{2}}\right\rangle\Bigr)=\left\langle   \vphantom{\bar{\tfrac{r}{2}}}\tfrac{r}{2}\Big| \tfrac{r}{2}\right\rangle\;\left\langle \bar{\tfrac{r}{2}}\Big|\bar{\tfrac{r}{2}}\right\rangle \neq 0.
\end{equation}
In a rank $r+1$ theory with $r+1>2$ and $r$ an odd integer the same problem occurs but on level $\mathbf{n}=1$: Such a theory possesses a level one representative $\tfrac{r+1}{2}\otimes\overline{\tfrac{r+1}{2}}$ with non-vanishing Shapovalov form. The crucial point is that though representatives of QSC equivalence classes are as per construction equivalent with respect to the action of $L_{0}$ and $\bar{L}_{0}$ in other respects their equivalence is not guaranteed. In every case is true: Representatives of QSC equivalence classes may be chosen such that observables in both formulations - QSC and KC - are identical. That is, even though $\mathcal{L}_{h\bar{h}\;QSC}\neq\mathcal{L}_{h\bar{h}\;KC}$ both representations are isomorphic and  $\left|h,\bar{h},\mathbf{n}\right\rangle_{QSC}$ is equivalent to $\left|h,\bar{h},\mathbf{n'}\right\rangle_{KC}$. Hence, at this point it is both impossible and unnecessary to finally rule on the question wether one method has to be prefered.

However, the known symplectic fermion realisation of the LCFT at $c=-2$ may be interpreted as a hint on the KC to be the more natural method. A symplectic fermion a la Zamolodchikov is a two-component fermionic field $\theta$ of spin zero \cite{GurFl97}. The stress energy tensor of its free theory is given by $T(z)=\frac{1}{2}\epsilon^{\alpha\beta}:\partial\theta_{\alpha}\,\partial\theta_{\beta}:(z)$.
The mode expansion of the component fields reads:
\begin{equation}
	\theta_{\alpha}=\sum\limits_{n\neq0}\theta_{\alpha,n}z^{-n}+\theta_{\alpha,0}\log(z)+\xi_{\alpha}.
\end{equation}
The $\xi$'s are Grassmann numbers and act as creation operators for the chiral logarithmic partner of the identity:
\begin{equation}	|h=0,k=1\rangle=-\frac{1}{2}\epsilon^{\alpha\beta}\,\xi_{\alpha}\,\xi_{\beta}\,|h=0,k=0\rangle=:\xi^{\alpha}\,\xi_{\beta}\,|0\rangle.
\end{equation}
Of course an analogue identity holds for the anti-chiral half. Therewith we can give explicit expressions for the sum of all states $\in \mathcal{R}_{0}\otimes\mathcal{R}_{\bar{0}}$ with equal level $n$: 
\begin{center}
\setlength{\unitlength}{.91pt}
\begin{picture}(70,70)(110,110)
 \put(-60,158) {$\mathbf{n=1}$}
  \put(11,158){$\bullet$}   \put(62,158){$\xi^{\alpha}\,\xi_{\beta}\,|0\rangle\otimes\bar{\xi^{\alpha}}\,\bar{\xi_{\beta}}\,|\bar{0}\rangle$} 
\put(160,158){$=\xi^{\alpha}\,\xi_{\beta}\,\bar{\xi^{\alpha}}\,\bar{\xi_{\beta}}\,|0\rangle\otimes\bar{0}\rangle$} 
  \put(17,153){\vector(1,-1){20}}
  \put(13,153){\vector(-1,-1){20}}  
  \put(-17,122){$\bullet$} 
  \put(38,122){$\bullet$} 
  \put(-60,122){$\mathbf{n=0}$}
\put(62,122){$\xi^{\alpha}\,\xi_{\beta}\,|0\rangle\otimes|\bar{0}\rangle+|0\rangle\otimes\bar{\xi^{\alpha}}\,\bar{\xi_{\beta}}\,|\bar{0}\rangle$} 
\put(217,122){$=\negmedspace\left(\xi^{\alpha}\,\xi_{\beta}+\bar{\xi^{\alpha}}\,\bar{\xi_{\beta}}\right)|0\rangle\otimes|\bar{0}\rangle$}
  \put(-15,118){\vector(1,-1){20}}
  \put(43,118){\vector(-1,-1){20}}      
  \put(11,87){$\bullet$} 
  \put(-60,87){$\mathbf{n=-1}$}
  \put(62,87){$|0\rangle\otimes|\bar{0}\rangle$}  
  \setlength{\unitlength}{1pt}
  \end{picture}
  \end{center}
  \vspace{12mm}
It is possible to choose a basis such that $\xi^{\alpha}=\bar{\xi^{\alpha}}$, i.\,e.\ the state $\left|\mathbf{n=1}\right\rangle$ vanishes due to the nilpotency of $\xi$. This coincides with the prediction of the kernel construction.
\section{Correlation functions}\label{cor_sec}
The prominent role that correlation functions play in CFTs results from two facts: On the one hand, they are related to observables and therefore represent a connection between theory and accessible experimental data. On the other hand, they are considered fundamental from a pure theoretical point of view: As shown in \cite{GabGod98}, a CFT is completely constituted if all correlation functions are known. Given the two- and three-point functions of the fundamental fields, all other amplitudes can actually be derived from these. Furthermore, the consistency conditions of all amplitudes can be traced back to those obeyed by the four-point functions.  

The calculation of correlation functions in LCFTs holds two major difficulties that do not arise in case of non-logarithmic CFTs. Both have their origin in the non-diagonal action of the generators of the chiral symmetry algebra. The off-diagonal contribution enters the global conformal Ward identities in the shape of an inhomogeneity and results in the aforementioned challenges:   

Firstly the identification of the generic structure of chiral correlation functions compatible to global conformal invariance is remarkably hindered. A hierarchical solution scheme for the inhomogeneous Ward identities allows to explore the texture of the subset of correlators which contain chiral fields residing in indecomposable representations whose irreducible subrepresentation corresponds to a proper primary field. For these cases, it is possible to fix the generic structure of $n$-point functions up to structure functions of $n-3$ $SL(2,\mathbb{C})$ invariant crossratios, however only within sets of other correlators. Possible extensions of the hierarchical solution scheme to pre-logarithmic fields and non-quasi-primaries are discussed in \cite{Flo01} and \cite{FloK03}.

Secondly, correlation functions of an LCFT do not generally factorise into chiral and anti-chiral parts. This is also an immediate consequence of the inhomogeneous Ward identities which is mirrored by the fact that only those correlators are factorisable that solve Ward identities with vanishing inhomogeneity. Gurarie pointed out that even amplitudes not explicitly involving logarithmic fields do not necessesarily fall in this category \cite{Gur93}. Two attemps have been made to adapt the knowledge about chiral correlation functions for non-chiral ones. One of us provided a rule of thumb, how to generalise known chiral sets of correlation functions to local sets by replacing all emerging variables $z_{i}-z_{j}=:z_{ij}$ by $\left|z_{ij}\right|^2$ \cite{Flo03}. This approach preserves the full generality of the chiral sets but obscures the interrelationship between chiral, anti-chiral and non-chiral amplitudes. Gaberdiel and Kausch suceeded in constructing a consistent set of amplitudes for the local theory at $c=-2$ \cite{GabK99}. As their course of action rests crucially upon model specific information like the operator product expansion (OPE) of the fundamental fields, it cannot be transfered to the generic case.   

Our proceeding will be as follows: We first recapitulate the nessessary assumptions under which generic chiral $n$-point functions can be calculated and briefly describe the hierachical solution scheme for these cases. Subsequently, we summarise the generic structure of the $n$-point functions found this way. In section \ref{loccor_sec} we will give a short proof for the statement that a non-chiral correlation function factorises if and only if the inhomogeneity of the Ward Identities vanishes for the chiral correlators it is composed of. Finally we present an ansatz built solely out of quantities enclosed in the chiral and anti-chiral sets of $n$-point correlators that allows the construction of local $n$-point amplitudes. 

Correlators of fields residing in the chiral (anti-chiral) half of an LCFT will be named chiral, anti-chiral respectively. Even though ``chiral'' intrinsically describes propagation properties of fields, we prefer this term to the adjunct holomorphic which is often chosen to indicate that a function only depends on the formal variable $z$ but not on $\bar{z}$. 

\subsection{Chiral correlation functions}\label{chircor_sec}
	
Correlation functions in LCFTs are invariant under the global conformal group. This stipulates the generic texture of $n$-point correlators up to structure functions of $n-3$ $SL(2,\mathbb{C})$ invariant crossratios that are a priori undetermined. Evaluating equation (\ref{Viraction_gl}) for $q=-1,0,1$ exhibits the inhomogeneous global conformal Ward identities (GCWIs): 
\begin{equation}\label{GWI_gl}
O_{q}\left\langle \ldots\right\rangle =-\sum^{n}_{i=1}z^{q}_{i}(q+1)\delta_{h_{i}}\left\langle \ldots\right\rangle, \qquad \mbox{for}\: q\in\left\{-1,0,1\right\}.
\end{equation}

From equation (\ref{GWI_gl}) follows immediately that correlation functions containing logarithmic fields cannot be determined independently: Due to the action of $\delta_{h_{i}}$, the generic structure of an $n$-point function including fields $\Psi_{(h_{i},k_{i})}(z_{i})$ can only be specified within a framework of other $n$-point functions containing fields $\Psi_{(h_{i},k_{i}-1)}(z_{i})$. For a given set of $n$ conformal weights $h_{i}$ with $r(h_{i})\negmedspace=\negmedspace r$ exists a hierarchy of $r^{n}$ $n$-point functions with $s=0,\ldots,n$ logarithmic fields displaying varying Jordan levels $k_i=1,\ldots,r-1$. The number of different correlation functions of a set is actually reduced, because a correlator is non-zero only, if the sum over the Jordan levels of the logarithmic fields it contains equals at minimum $r-1$ \cite{Flo01}:
\begin{equation}\label{Anfbed_gl}
	\left\langle k_{1}\ldots k_{n}\right\rangle\neq0 \ \Leftrightarrow  \; K:=\sum^{n}_{i=1}k_{i}\geq r-1.
\end{equation}
To improve lucidity we shortened the naming of the correlators in equation (\ref{Anfbed_gl}):
\begin{equation*}
\left\langle k_{1}\ldots k_{n}\right\rangle:=\left\langle\Psi_{(h_{1},k_{1})}(z_{1})\ldots\Psi_{(h_{n},k_{n})}(z_{n})\right\rangle.
\end{equation*}  
 
The identity (\ref{Anfbed_gl}) serves as starting point for a recursive construction of solutions of the GCWIs (\ref{GWI_gl}). For total Jordan level $K=r-1$, the GCWIs are homogeneous and can be solved as known from ordinary CFT. Successive increase of the total Jordan level yields differential equations for $\left\langle k_{1}\ldots k_{n}\right\rangle$ with the inhomogeneity $\sum_{i}\delta_{h_{i}}\left\langle k_{1}\ldots k_{n}\right\rangle$ determined in foregoing steps of the recursion.  

One finds that $n$-point correlators which contain fields of rank $r$ Jordan cells possess the generic form: 
\begin{equation}\label{GF_gl}
\left\langle k_{1}\ldots k_{n}\right\rangle=\prod_{i<j}(z_{i}-z_{j})^{\mu_{ij}}\sum_{G=0}^{l^{max}}F^{G}_{\left\{q\right\}}(x_{a})\cdot P_{G}(l_{mn}),\qquad l_{mn}:= \ln\left(z_{mn}\right)
\end{equation}
where $P_{G}$ denotes a sum over monomials of degree $G$: 
\begin{equation}\label{Poly_gl}
P_{G}=\sum_{\alpha|g(\alpha)=G}c_{\alpha}\prod^{j}_{i=1}(l_{m_{i}n_{i}})^{g_{\alpha_{i}}}=:\sum_{\alpha}c_{\alpha}p_{\alpha},  \quad \sum^{j}_{i=1}g_{\alpha_{i}}=g(\alpha). 
\end{equation}
The constraint of global conformal invariance (\ref{GWI_gl}) connects a coefficient $c_{\alpha}$ multiplying a monomial $p_{\alpha}$ in a correlator A to the coefficient $c_{\beta}$ which multiplies a monomial $p_{\beta}$ in a correlator $\sum_{i}\delta_{h_{i}}A$, with $p_{\beta}$ being the image of $p_{\alpha}$ under the action of $O_{0}$. Four-point or higher correlation functions that exhibit logarithmic fields at every vertex may feature polynomials $K_{G}\subset P_{G}$, whose multiplicities are special in the following sense: They are not cross-linked to any coefficients in other $n$-point functions of the set. Linkage to correlators with lower total Jordan level is canceled if $K_{G}$ resides in the kernel of the operator $O:=(O_{0},O_{1})$ \cite{FloK03}. No linkage to higher correlators of the set has to be required seperatly. It follows that kernel terms $K_{G}$ may only arise in the highest correlator of a set, i.\,e.\ in the correlator where all inserted fields are of maximum Jordan level $k_n=r-1$. 

The occurring monomials $p_{\alpha}$ are subject to selection rules \cite{Flo03}: Logarithms in the correlators stem from contractions of logarithmic fields. Hence, only such logarithms may arise whose indices refer to positions of fields with Jordan level $k\geq1$ within a correlator. Two more restrictions rule the logarithmic terms: 
\newcounter{Lcount}
\begin{list}{[S\arabic{Lcount}]}{\usecounter{Lcount}\setlength{\rightmargin}{\leftmargin}}
\item The total logarithmic degree G in (\ref{GF_gl}) is bounded above as follows:
	\[  
  G\leq K-r+1=:l^{max}.
\]
\item Each index $m_{i}$ may arise at most $r-1$ times within one monomial. 
\end{list}
  
Let $F^{G}_{\left\{q\right\}}(x_{a})$ denominate a family of functions, which solely depend on $n-3$ anharmonic ratios $x_{a}$. The subscript $\left\{q\right\}$ denotes a set of $n$ indices $q_{i}$ which take integer values between zero and $k_{i}$:
	\[F^{G}_{\left\{q\right\}}:=F^{G}_{q_{1}q_{2}\ldots q_{n}}, \qquad q_{i}\in\left\{0,1 \ldots k_{i}\right\}.
\]
For fixed superscript index $G$, only those combinations ${\left\{q\right\}}$ emerge that fulfil
	\[\sum _{i}q_{i}+G=K.
\]
Due to the cluster decomposition property all structure functions that satisfy $\sum q_{i}=r\negmedspace-\negmedspace1$ have to be identified \cite{FloK03} and will be refered to as $F^{l^{max}}$. For $n\leq3$ $F^{G}_{\left\{q\right\}}$ does not depend on the values $q_{i}$ but only on $\sum _{i}q_{i}$ \cite{Flo01}. 

The structure functions $F_{\left\{{q}\right\}}(x)$ may be decomposed in conformal blocks $\mathcal{F}^{i}_{\left\{q\right\}}(x)$ which represent the internal propagators: 
	\[F_{\left\{q\right\}}(x)=\sum_{i}\mathcal{F}^{i}_{\left\{q\right\}}(x).
\]
During the main part of the paper this decomposition will not play any role for the presented argumentation. To keep things simple we will abstain from making it explicitly where it is not necessary.

The exponents $\mu_{ij}$ in equation (\ref{GF_gl}) solve 
\[\sum_{i\neq j}\mu_{ij}=-2h_{j}.
\]
 
So far we have stated properties of chiral correlation functions only. It is clear that analogues propositions hold for anti-chiral correlators.

\subsection{Assembling local amplitudes from the chiral sets} \label{loccor_sec}
     
We can now bring the original query into sharper focus. Non-chiral amplitudes shall be obtained by multiplying suitable chiral and anti-chiral amplitudes. Therefore we have to revisit the constraints of locality (\ref{LC_gl}). As in the frame of this paper only the case $h_{i}=\bar{h_{i}}$ is considered, the first condition does not cause concern. The second condition shall, for our purpose, be restated as constraint on correlation functions. Using equation (\ref{GWI_gl}), one finds:	\[-\sum^{n}_{i=1}S_{i}\left\langle\Psi_{1}\ldots\Psi_{n}\right\rangle=\left(O_{0}-\bar{O_{0}}\right)\left\langle\Psi_{1}\ldots\Psi_{n}\right\rangle=0.
\]
In non-logarithmic conformal field theories, the non-chiral amplitudes $\mathbf{A}$ can be achieved by multiplying the chiral and anti-chiral amplitudes $A$ and $\bar{A}$:
\begin{equation}\label{AbarA_gl}\mathbf{A}=A\bar{A}.
\end{equation}
In LCFTs, factorisation is contradictory to locality constraints except for correlators satisfying homogeneous Ward identities: 
\begin{align}\label{LocAn_gl}
(O_{0}-\bar{O}_{0})\mathbf{A}& 
 =(O_{0}A)\bar{A}-A(\bar{O}_{0}\bar{A})\\
&=\negthinspace-\negthinspace\left(\sum_{i}\delta_{h_{i}}A\right)\bar{A}+A\left(\sum_{i}\delta_{\bar{h}_{i}}\bar{A}\right)\negmedspace=0\;\Leftrightarrow\; \sum_{i}\delta_{h_{i}}A=\sum_{i}\delta_{\bar{h}_{i}}\bar{A}\equiv0 \nonumber
\end{align}
where the first identity follows from the fact, that the operators $O_{0}$ and $\bar{O}_{0}$ act as derivatives with respect to $z$ ($\bar{z}$ respectively) on the function space, i.\,e.\ $\bar{O}_{0}$ acting on the chiral amplitude does not yield a contribution and vice versa. The second identity arises out of the chiral amplitudes satisfying the GCWIs. The last step is based on the fact, that the maximum logarithmic degree of $\delta_{h_{i}}A$ is reduced by one compared to the maximum logarithmic degree of $A$. 

The given argumentation is not affected if the conformal block decomposition of the amplitudes is taken into account, i.\,e.\ if equation \eqref{AbarA_gl} is substituted by
	\[\mathbf{A}=\sum_{qp}\mathcal{X}_{qp}A_q\bar{A}_p,
\]    
where $A_q$ denotes the contribution of a conformal block $q$ to $A$. As the conformal blocks do only depend on the crossratios, adjustment of their linear combination can not cancel the mismatch of logarithmic powers in equation \eqref{LocAn_gl}.

Subsequently, we present an ansatz that admits the construction of generic local $n$-point functions out of the known chiral correlators:  
\begin{equation}\label{ansatznc_gl}
\left\langle\mathbf{k_{1}}\ldots \mathbf{k_{n}}\right\rangle=\left\langle\vphantom{\overline{k_{n}}}k_{1}\ldots k_{n}\right\rangle\left\langle \overline{k_{1}\ldots k_{n}}\right\rangle\vert_{\mbox{selection rules}}+GOL,  
\end{equation}
where $GOL$ stands for guarantor of locality and lives up to its name by providing the desired behavior of $\left\langle\mathbf{k_{1}}\ldots \mathbf{k_{n}}\right\rangle$ under the action of $O_{0}-\bar{O}_{0}$. 
This fixes $(O_{0}\nolinebreak-\nolinebreak \bar{O}_{0})\nolinebreak \;GOL$ as follows:
\begin{equation} \label{O-OGOL_gl}
(O_{0}-\bar{O}_{0})GOL=\sum^{n}_{i=1}\biggl[\Bigl(\delta_{h_{i}}\left\langle\vphantom{\overline{k_{n}}} k_{1}\ldots k_{n}\right\rangle\Bigr)\left\langle \overline{k_{1}\ldots k_{n}}\right\rangle-\left\langle\vphantom{\overline{k_{n}}} k_{1}\ldots k_{n}\right\rangle\Bigl(\delta_{\bar{h}_{i}}\left\langle \overline{k_{1}\ldots k_{n}}\right\rangle\Bigr)\biggr].	
\end{equation}
The contribution of 
\begin{equation*}
	\left\langle \vphantom{\overline{k_{n}}}k_{1}\ldots k_{n}\right\rangle\negthickspace\left\langle
\overline{k_{1}\ldots k_{n}}\right\rangle\negthickspace=\negthickspace\prod_{i<j}\left|z_{ij}\right|^{2\mu_{ij}}\!\negthickspace \left[\sum_{g=0}^{l^{max}}F^{G}_{\left\{q\right\}}(x)P_{G}(l_{m_{s}n_{s}})\right]\negthickspace\left[\sum_{g=0}^{l^{max}}\bar{F}^{\bar{G}}_{\left\{\bar{q}\right\}}(\bar{x})\bar{P}_{\bar{G}}(\overline{l_{m_{r}n_{r}}})\right]
\end{equation*}
in equation (\ref{ansatznc_gl}) is constricted to terms satifying selection rules. The selection rules for the arising logarithmic terms in chiral correlators have been resumed in section \ref{chircor_sec}. Generalising them to the non-chiral case is straightforward: The highest logarithmic degree to appear in a correlator was shown to depend on its total Jordan level and the rank of the theory, [$\mbox{S1}$]. As demonstrated in section \nolinebreak\ref{content_sec} the rank of the non-chiral theory equals the rank of the chiral halves it is composed of. Furthermore, as per construction the total Jordan level of the left hand side of equation (\ref{ansatznc_gl}) equals the total Jordan level of the chiral correlators $\left\langle\vphantom{\overline{k_{n}}} k_{1}\ldots k_{n}\right\rangle$ and $\left\langle \overline{k_{1}\ldots k_{n}}\right\rangle$, i.\,e.\:
 \[g+\bar g\leq l^{max},\qquad l^{max}_{non-chiral}=l^{max}_{chiral}=l^{max}_{anti-chiral}.
\]
The highest multiplicity for one index to appear within a monomial was stated to solely depend on the rank of the theory, [$\mbox{S2}$]. Thus, according to the aforementioned reasoning, it can be adopted from the chiral case as its stands.  
 
Therewith the left hand side of the ansatz (\ref{ansatznc_gl}) can in principle be calculated. For that purpose, we have to expand the product of the chiral correlators, implement [\mbox{S1}] and [\mbox{S2}] and add an expansion of $GOL$. The latter  is given as follows:
\begin{equation}\label{GOL1_gl}
GOL=\negthickspace\prod_{i<j}\left|z_{ij}\right|^{2\mu_{ij}}\sum_{G=0}^{l^{max}}\sum_{\alpha,\beta}c_{\alpha\beta}\prod^{t}_{s=1}\prod^{u}_{r=1}(l_{m_{s}n_{s}})^{g_{\alpha_{s}}}(\overline{l_{m_{r}n_{r}}})^{\bar{g}_{\beta_{r}}}, \;  \sum^{t}_{s=1}\sum^{u}_{r=1}(g_{\alpha_{s}}+\bar{g}_{\beta_{r}})\negmedspace=\negmedspace G.
\end{equation}
It is clear that the coefficients $c_{\alpha\beta}$ depend on the chiral and anti-chiral structure functions, i.\,e.\ $c_{\alpha\beta}=c_{\alpha\beta}\left(F_{\left\{q\right\}},\bar{F}_{\left\{\bar{q}\right\}}\right)$.
For further convenience, we will subsequently use the naming convention introduced in equation (\ref{Poly_gl}):  
\[p_{\alpha}:=\prod^{t}_{s=1}(l_{m_{s}n_{s}})^{g_{\alpha_{s}}}, \qquad \sum^{t}_{s=1}g_{\alpha_{s}}=g(\alpha)=:g_{\alpha}.
\]
Herewith the expansion of $GOL$ (\ref{GOL1_gl}) reduces to the form:  
\begin{equation}
GOL=\prod_{i<j}\left|z_{ij}\right|^{2\mu_{ij}}\sum_{G=0}^{l^{max}}\sum_{\alpha,\beta}c_{\alpha\beta}p_{\alpha}\bar{p}_{\beta}.
\end{equation}

Of course, the logarithmic terms of $GOL$ are subordinated to $[\mbox{S1}]$ and $[\mbox{S2}]$, too. Further restrictions on the structure of $GOL$ arise from its claimed behavior under the action of $(O_{0}-\bar{O_{0}})$: Equation (\ref{O-OGOL_gl}) changes sign under complex conjugation. It follows that
\begin{equation}\label{structureGOL_gl}
GOL=\overline{GOL}\quad\Rightarrow\quad c_{\alpha\beta}=\overline{c_{\beta\alpha}},\ \: \mbox{in particular}\;\:c_{\alpha\alpha}\in\mathbb{R}.
\end{equation}
 
In addition the coefficients $c_{\alpha\beta}$ are coupled to linear combinations of the structure functions $F^{G}_{\left\{q\right\}}(x)$ and $\bar{F}^{\bar{G}}_{\left\{\bar{q}\right\}}(\bar{x})$ by two sets of constraints. The first set arises out of the condition (\ref{O-OGOL_gl}). The second emanates from the logarithmic identities governing the assembly of local monomials: Monodromy invariance of equation (\ref{ansatznc_gl}) enforces arguments of emerging logarithms to be real, i.\,e.\ logarithmic terms have to be of the shape 
\begin{equation}\label{locmonomials_gl}
\begin{split}
	\mathbf{p}_{\alpha}(\left|l_{m_{s}n_{s}}\right|^2):&=\prod^{t}_{s=1}(\left|l_{m_{s}n_{s}}\right|^2)^{g_{s}}\\
	&=\prod^{t}_{s=1}\left(l_{m_{s}n_{s}}+\overline{l_{m_{s}n_{s}}}\right)^{g_{s}}\\
&=\prod^{t}_{s=1}\left(\sum^{g_{s}}_{i=0}\binom{g_{s}}{i}\left(\vphantom{\overline{l}}l_{m_{s}n_{s}}\right)^{i}\left(\overline{l_{m_{s}n_{s}}} \right)^{g_{s}-i}\right).
\end{split}
\end{equation}
Hence, coefficients of terms $\prod^{t}_{s=1}\left(\vphantom{\overline{l}}l_{m_{s}n_{s}}\right)^{i}\left(\overline{l_{m_{s}n_{s}}} \right)^{g_{s}-i}$ with $0\leq i\leq g_{s}$ are fixed up to an overall factor. These coefficients are proportional to $F^{G}_{\left\{q\right\}}(x)\bar{F}^{\bar{G'}}_{\left\{\bar{q'}\right\}}(\bar{x})$ if a monomial stems from $\left\langle\vphantom{\overline{k_{n}}}k_{1}\ldots k_{n}\right\rangle\left\langle \overline{k_{1}\ldots k_{n}}\right\rangle$ or else given by $c_{\alpha\beta}$, which establishes the aforementioned coupling. Recalculating one finds, that every solution of the resulting set of constraints solves equation (\ref{O-OGOL_gl}). 

Without loss of generality the free choice of an overall factor of equation (\ref{locmonomials_gl}) can be absorbed in the factors of contributions $\prod^{t}_{s=1}\left(\vphantom{\overline{l}}l_{m_{s}n_{s}}\right)^{g_{s}}\left(\overline{l_{m_{s}n_{s}}} \right)^{0}$. This lightens our ansatz (\ref{ansatznc_gl}) to
\begin{equation}\label{solutionnc_gl}
	\left\langle\mathbf{k_{1}}\ldots \mathbf{k_{n}}\right\rangle=\prod_{i<j}\left|z_{ij}\right|^{2\mu_{ij}}\sum_{G=0}^{l^{max}}\sum_{\alpha}\underbrace{\left(\bar{F}_{\left\{\bar{q}'\right\}}^{\bar{0}} F^{G}_{\left\{q\right\}}c_{\alpha}+c_{\alpha0}\right)}_{=:\mathbf{C}_{\alpha}}\mathbf{p}_{\alpha}.
\end{equation}

Within this generic approach the monodromy properties of the local correlators cannot be completely explored. The structure functions $\bar{F}_{\left\{\bar{q}'\right\}}(\bar{x})$ and $F_{\left\{q\right\}}(x)$ in equation \eqref{solutionnc_gl} are linear combinations of conformal blocks: 
	\[\bar{F}_{\left\{\bar{q}'\right\}}(\bar{x})=\sum_{i}\bar{\mathcal{F}}^{i}_{\left\{\bar{q}'\right\}}(\bar{x}),\quad\, F_{\left\{q\right\}}(x)=\sum_{j}\mathcal{F}^{j}_{\left\{q\right\}}(x),\quad\, \bar{F}_{\left\{\bar{q}'\right\}}F_{\left\{q\right\}}=\sum_{ij}\mathcal{X}_{ij}\bar{\mathcal{F}}^{i}_{\left\{\bar{q}'\right\}}\mathcal{F}^{j}_{\left\{q\right\}}.
\]
Enforcing monodromy invariance of the amplitude \eqref{solutionnc_gl} determines the coefficients $\mathcal{X}_{ij}$, a task that cannot be performed within the generality aimed here.  

It is worth pointing out that the claims asserted so far suffice to guarantee all coefficients $\mathbf{C}_{\alpha}$ being real: The constraint (\ref{locmonomials_gl}) in particular demands that the $i=0$ term and the $i=g_s$ term arise with the same multiplicity. The coefficient of the former can easily be shown to equal $\bar{F}_{\left\{\bar{q}\right\}}^{\bar{G}} F^{0}_{\left\{q'\right\}}c_{\alpha}+c_{0\alpha}$. Using equation (\ref{structureGOL_gl}) we can conclude:  
\[\bar{F}_{\left\{\bar{q}\right\}}^{\bar{G}} F^{0}_{\left\{q'\right\}}c_{\alpha}+c_{0\alpha}=\overline{\mathbf{C}_{\alpha}}=\mathbf{C}_{\alpha}\quad\Rightarrow\mathbf{C}_{\alpha}\in \mathbb{R}.
\]  

Reimplementing the GCWIs establishes dependencies between coefficients $\mathbf{C}_{\alpha}$ of different local correlators of a set. The occurring cross-linkage of coefficients is in perfect analogy to the chiral case: $\mathbf{C}_{\alpha}$ emerging in a correlator $\mathbf{A}$ is connected to $\mathbf{C}_{\beta}$ in a correlator $\sum_{i}\delta_{h_{i}}\mathbf{A}=\sum_{i}\delta_{\bar{h}_{i}}\mathbf{A}$ if $\mathbf{p}_{\beta}$ is the image of $\mathbf{p}_{\alpha}$ under the action of $O_{0}$ as well as under the action of $\bar{O}_{0}$. It follows that the obtained solution for a generic chiral set of correlation functions, obeying locality and global conformal invariance is not unique. 

The number of degrees of freedom a local correlator possess equals the number of those the corresponding chiral correlator shows. This matches with the predictions of the substitution method \cite{Flo03}. Anyhow it is  astonishing as one could have expected that implementing the condition of locality would confine the number of free parameters. Investigating duality \cite{DiFr97,GSW87} of the obtained generic correlation functions yields that this constraint too does not reduce the number of free parameters any further.

\subsection{Discussion and summary}

\subsubsection*{Non-uniqueness of the solution}
Generic local sets of correlators cannot uniquely be determined by virtue of locality, duality and global conformal invariance. The most general form of $GOL$ preserves the number of free parameters a set of correlation functions exhibits. The minimal choice for $GOL$ fulfilling the constraints (\ref{O-OGOL_gl}) and (\ref{locmonomials_gl}) is build up of mixed terms only, i.\,e.\ it does not exhibit any contributions $p_{\alpha}\bar{p}_{\beta}$ with $g_\alpha=0$ or $\bar{g}_\beta=0$. As a consequence, $c_{\alpha0}$ in equation (\ref{solutionnc_gl}) vanishes identically for all $\alpha$. In this case, reimplementing the GCWIs fixes all but one conformal block of a set. We are confronted with a situation very similar to non-logarithmic CFT: Although in the LCFT case a whole hierarchy of $n$-point functions emanates from a set of n conformal weights $h_i=\bar{h}_i$ the structure of each correlator is fixed up to one shared structure function $F$ which solely depends on $n-3$ crossratios and the conformal weights $h_i$ but not on $\mathbf{k_{i}}$.    
     
\subsubsection*{Consistency check}
Our method allows us to connect the results of \cite{FloK03} to the results of \cite{GabK99}.
According to \cite{FloK03} the rank two chiral set of four-point functions with $h_{i}=0\,\forall i$ reads
\begin{subequations}\label{set1_gl}
\begin{align}
  \left\langle1000\right\rangle = & F_{1000} \; , \label{1first}\\ 
  \left\langle1100\right\rangle = & \mathcal{P}_{S_2} \Big\{ \tfrac{1}{2}  F_{1100} - l_{12} F_{1000} \Big\} \; , \label{1second}\\
  \left\langle1110\right\rangle = & \mathcal{P}_{S_3} \Big\{ \tfrac{1}{6}  F_{1110}+(\tfrac{1}{2} l_{23}-l_{12}) F_{0110}+ \big[ l_{12}l_{23}-\tfrac{1}{2} l^{2}_{12}\big] F_{1000} \Big\} \; , \label{1third}\\
 \left\langle1111\right\rangle = & \mathcal{P}_{S_4} \Big\{  \tfrac{1}{24}  F_{1111} + (\tfrac{1}{6} l_{23}-\tfrac{1}{3}l_{12})F_{0111} + \label{1fourth}\\
  & \phantom{\mathcal{P}_{S_4} \Big\{} \big[ \tfrac{1}{2}  (l_{14}l_{23} - l_{12}l_{34})+( l_{12}l_{23} -\tfrac{1}{2}l_{24}l_{23} ) - \tfrac{1}{4} l^{2}_{12} \big] F_{0011} + \nonumber\\
  & \phantom{\mathcal{P}_{S_4} \Big\{} \big[ \tfrac{1}{2} l^{2}_{12}l_{34}+\tfrac{1}{3} l_{12}l_{23}l_{13}- l_{12}l_{23}l_{34} \big] F_{1000} \big\}  \; .\nonumber
\end{align} 
\end{subequations}
Missing correlators of the set can be obtained by permutation of the inserted fields. Let $\mathcal{P}_{S_x}$ denote the sum over all permutations of indices generated by the group ${S_x}$. As all four fields posses the same conformal weight $h_{i}=h_{j}$ it follows from the associativity of the fusionalgebra that  
\begin{equation}\label{perminv_gl}
 F_{q_{1}q_{2}q_{3}q_{4}}(x)= F_{q_{\sigma(1)}q_{\sigma(2)}q_{\sigma(3)}q_{\sigma(4)}}(x) \qquad \forall \sigma \in S_{4},
\end{equation}
i.\,e.\ $\mathcal{P}_{S_x}$ does only affect the indices of logarithmic terms $l_{ij}$. The symmetry (\ref{perminv_gl}) admits to identify all $F_{\mathbf{q}}$ with equal $\sum q_{i}=p$ and to abbreviate them by the $p$-th capital letter of the latin alphabet. As $\Psi_{h=0, k=0}$ denotes the identity field the amplitudes (\ref{set1_gl}a-c) are identical to the one- two- and three-point functions of the field $\Psi_{h=0,\, k=1}$. Consequentially, in the following expressions, $A$, $B$ and $C$ are constant, $D=D(x)$. We can now arrange the equations \eqref{set1_gl} more clearly:
\begin{subequations}\label{set2_gl}
\begin{align}
  \left\langle1000\right\rangle = & A \; , \label{2first}\\ 
  \left\langle1100\right\rangle = & B-2A\left(l_{12}\right) \; , \label{2second}\\
  \left\langle1110\right\rangle = & C-B\left(l_{12}+l_{13}+l_{23}\right)+\label{2third}\\
  &A\big[ 2\left(l_{12}l_{23}+l_{13}l_{23}+l_{12}l_{13}\right)-\left(l^{2}_{12}+l^{2}_{13}+l^{2}_{23}\right)\big]  \; ,\nonumber\\ 
 \left\langle1111\right\rangle = & D-\frac{2}{3}C\ast\negthickspace-\negthickspace\ast+2B\big[(2\kappa+1)\ast\negthickspace-\negthickspace\ast\ast\negthickspace-\negthickspace\ast-\kappa\ast\negthickspace-\negthickspace\ast\negthickspace-\negthickspace\ast+\kappa\ast\negthickspace=\negthickspace\ast\big]+\label{2fourth}\\
 & 2A\big[\ast\negthickspace=\negthickspace\ast\ast\negthickspace-\negthickspace\ast+ \vtop{\ialign{#\crcr ${\ast\negthickspace-\negthickspace\ast\negthickspace-\negthickspace\ast}$\crcr \noalign{\nointerlineskip} $\kern3pt\vrule height4pt\hrulefill\vrule height4pt\kern3pt$\crcr}}-\ast\negthickspace-\negthickspace\ast\negthickspace-\negthickspace\ast\negthickspace-\negthickspace\ast\big]   \; .\nonumber
\end{align} 
\end{subequations} 
The correlation function \eqref{2fourth} exhibits a kernel term $K_G$ whose multiplicity $\kappa$ is not fixed by global conformal invariance. Moreover in support of clarity we chose a graphical notation to depict the logarithmic terms in this correlator. Reading the diagrams is straightforward: Each position denotes a variable indexvalue. The symbol filling a position designates the variable which carries the index. We'll use $\ast$ for $z$ and $\triangle$ for $\left|z\right|^{2}$. A line between two symbols of the same kind indicates the logarithm of the symbollically represented variable. Each graph stands for the sum over all those identifications of variable indexvalues with numerical indexvalues which provide expressions that are not equivalent, e.\,g.\
	\[ \triangle\negthickspace-\negthickspace\triangle\triangle\negthickspace-\negthickspace\triangle=\mathbf{l}_{12}\mathbf{l}_{34}+\mathbf{l}_{13}\mathbf{l}_{24}+\mathbf{l}_{14}\mathbf{l}_{23},\qquad \mathbf{l}_{ij}:=\ln(\left|z_{ij}\right|^{2}).
\]

Combining local amplitudes according to the method described above yields: 
\begin{subequations}\label{set3_gl}
\begin{align}
  \left\langle\mathbf{1000}\right\rangle\! = & \mathbf{C}_{AA} \; , \label{3first}\\ 
  \left\langle\mathbf{1100}\right\rangle\! = & \mathbf{C}_{BB}-2\mathbf{C}_{BA}\left(\mathbf{l}_{12}\right) \; , \label{3second}\\
  \left\langle\mathbf{1110}\right\rangle\! = & \mathbf{C}_{CC}-\mathbf{C}_{CB}\left(\mathbf{l}_{12}+\mathbf{l}_{13}+\mathbf{l}_{23}\right)+\label{3third}\\
  &\mathbf{C}_{CA}\big[ 2\left(\mathbf{l}_{12}\mathbf{l}_{23}+\mathbf{l}_{13}\mathbf{l}_{23}+\mathbf{l}_{12}\mathbf{l}_{13}\right)-\left(\mathbf{l}^{2}_{12}+\mathbf{l}^{2}_{13}+\mathbf{l}^{2}_{23}\right)\big]  \; ,\nonumber\\
 \left\langle\mathbf{1111}\right\rangle\! = & \mathbf{C}_{DD}-\frac{2}{3}\mathbf{C}_{DC}\triangle\negthickspace-\negthickspace\triangle+2\mathbf{C}_{DB}\big[(2\lambda+1)\triangle\negthickspace-\negthickspace\triangle\triangle\negthickspace-\negthickspace\triangle-\lambda\triangle\negthickspace-\negthickspace\triangle\negthickspace-\negthickspace\triangle+\nonumber\\
 &\lambda\triangle\negthickspace=\negthickspace\triangle\big]+ 2\mathbf{C}_{DA}\big[\triangle\negthickspace=\negthickspace\triangle\triangle\negthickspace-\negthickspace\triangle+ \vtop{\ialign{#\crcr ${\triangle\negthickspace-\negthickspace\triangle\negthickspace-\negthickspace\triangle}$\crcr \noalign{\nointerlineskip} $\kern3pt\vrule height4pt\hrulefill\vrule height4pt\kern3pt$\crcr}}-\triangle\negthickspace-\negthickspace\triangle\negthickspace-\negthickspace\triangle\negthickspace-\negthickspace\triangle\big]   \; ,\label{3fourth}
\end{align} 
\end{subequations}
where $\mathbf{C}_{XY}=\bar{X}Y+c_{xy}$ with $c_{xy}$ being a complex number corresponding to $c_{\alpha0}$ in formula (\ref{solutionnc_gl}). The reimplementation of the GCWIs enforces the identification all coefficients $\mathbf{C}_{XY}$ with same second index: $\mathbf{C}_{XY}\equiv\mathbf{C}_{Y}\ \forall\; X$. It is worth noticing that in the course of assembling local amplitudes the arbitrary multiplicities $\kappa$ and $\bar{\kappa}$ of the chiral kernel term are set to zero. The kernel term $\mathbf{\lambda K}_G=\lambda\left(2\,\triangle\negthickspace-\negthickspace\triangle\triangle\negthickspace-\negthickspace\triangle\,-\,\triangle\negthickspace-\negthickspace\triangle\negthickspace-\negthickspace\triangle\,+\,\triangle\negthickspace=\negthickspace\triangle\right)$ in the correlator \eqref{3fourth} is exclusively composed of contributions of $GOL$. 

We can now resume: The equations \eqref{set3_gl} constitute a generic set of correlators containing fields from a rank two reducible representation with $h=\bar{h}=0$. One explicitly known representative of this case is given through the set of $n$-\nolinebreak[4]point functions $\left\langle\prod_{n}\Psi_{h_n=0,\bar{h}_n=0,\mathbf{k_n=1}}\right\rangle$, $n=1,\ldots4$ of the local LCFT at $c=-2$ \cite{GabK99}. Except for an overall sign of the correlator (\ref{3fourth}) both sets - the generic and the concrete - are consistent: For the local triplet theory, the undetermined coefficients of the generic set take the values 
\begin{align}
\mathbf{C}_A&=\mathcal{C}_0=1,& \mathbf{C}_B&=-8\ln(2),\nonumber\\ \mathbf{C}_C&=48\ln^2(2),& \mathbf{C}_D&=-256\ln^3(2).\nonumber 
\end{align}
Furthermore the kernel multiplicity $\lambda$ is fixed to $-\frac{1}{2}$ in virtue of the operator product expansion.

\section{Conclusions}

Exploiting the conformal symmetry allows first steps towards the construction of generic local LCFTs. The generality of our approach is confined by the following assumptions: Indecomposable representations are considered only with respect to the Virasoro zero mode. Irreducible subrepresentations are assumed to correspond to proper primaries, logarithmic partners to quasi-primaries. Furthermore only the diagonal case $h_i=\bar{h}_i$ is regarded.  

For future work, generalising our findings by relaxing these assumptions would be an interesting task: As mentioned before the scope of section \ref{cor_sec} can possibly be extended to pre-logarithmic fields and non-quasi-primaries \cite{Flo01, FloK03}. Remarkably, it seems that even abdicating the $h_i=\bar{h}_i$ condition does not seriously damage the proposed method for the construction of generic local correlation functions out of generic chiral ones. 

The assumption that irreducible subrepresentations correspond to proper primary
fields, as made in our work, is more of a technical nature to simplify the
deductions. Furthermore, it is the only case for which the general form of
four-point functions has been explicitly computed. However, the algorithm
presented in \cite{FloK03} can easily be adapted to deal with a more
general setting. Essentially, this amounts into setting a less simple
initial condition. We stress that this is of relevance as the irreducible
subrepresentation with weight $h=5/8$ in the augmented $c=c_{2,3}$ model
provides a counter-example to our assumption \cite{EF06}.

The assumption of quasi-primarity of logarithmic partner fields is the
only assumption we really need and which certainly is violated in many
cases. For example, representations of all types but type A in the notation
of \cite{EF06} do violate the assumption. It is also violated in
general in the $c_{p,1}$ models for $p>2$. Even the standard example,
the $c_{2,1}=-2$ model, contains one reducible but indecomposable 
representation, where this assumption is broken, ${\cal R}_1$.
However, as argued in \cite{Flo01}, a fermionic zero mode content can
be defined for the $c=-2$ theory. It can then be shown that the part of
the field which violates quasi-primarity differs in its fermionic zero mode
content by one modulo two. As a consequence, this part does not affect
correlation functions, see \cite{Flo01} for details. As all
$c_{p,1}$ models enjoy structurally similar fermionic sum representations
of their partition function and characters as for the $c=-2$ model
\cite{FGK07}, in particular
a symplectic structure, it seems very plausible that one can define a
fermionic zero mode content for all $c_{p,1}$ models. This might indicate
that the condition of quasi-primarity could be relaxed for these models
in the same way as for the $c=-2$ model. However, we do not know whether
this might generalize to the augmented minimal models. 

On the other hand,
the representation ${\cal R}_1$ in the $c=-2$ model actually is a 
representation of a larger symmetry than the Virasoro algebra, namely a 
${\cal W}$-algebra representation of the triplet algebra.
In this particular case, it turns out that the non-quasi-primary state,
let it call us $|\phi\rangle$, has the property that $L_1|\phi\rangle
\propto W^0_1|\phi\rangle$. Thus, this degeneracy in the space of states
can be used to eliminate contributions from the non-quasi-primarity of
$|\phi\rangle$.
Thus, the
problem of non-quasi-primarity occurs in the $c=-2$ model on the level
of ${\cal W}$-descendants of ${\cal W}$-primaries and their 
${\cal W}$-log-partners. As correlation functions of descendant fields are
uniquely determined through the ${\cal W}$-symmetry, as soon as the 
correlation functions
of the basic fields are known, one may argue that the problem of
non-quasi-primarity is -- on the level of the representation theory of 
the maximally extended symmetry algebra -- not that relevant. The same
argument should hold for the structurally very similar $c_{p,1}$ models
with their triplet algebras \cite{CF06}. As the augmented minimal models also
possess extended symmetry algebras, we may hope that non-quasi-primarity
does not play a role for the basic fields in the full 
${\cal W}$-algebra representations.

Nevertheless,
a matter of particular interest would be to investigate the local space of states for a theory whose chiral halves violate the assumption of quasi-primarity of the logarithmic partner fields. In case of chiral Jordan cells which contain at least one logarithmic field $\Psi_{h,k}$ ($\bar{\Psi}_{\bar{h},\bar{k}}$ respectively) with the property
	\[\exists\ n>0:\ \lim_{z\rightarrow0} L_n\,\Psi_{h,k}\,(z)\left|\,0\right\rangle\neq0 
\]
the equivalence of QSC and KC is broken. This shall be illustrated by means of a rank three non-chiral representation where the chiral Jordan level one fields are not quasi-primary.    
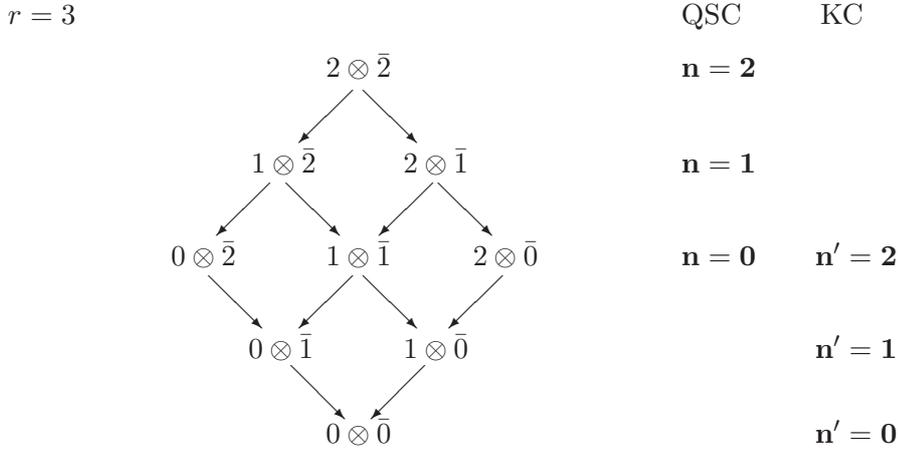
\begin{center}
\begin{picture}(220,170)(-40,20)
\put(136,178){QSC}
\put(188,178){KC} 
 \put(-116,178) {$r=3$}
   \put(136,158){$\mathbf{n=2}$}
  \put(3,158){$2\otimes\bar{2}$} 
  \put(17,153){\vector(1,-1){20}}
  \put(13,153){\vector(-1,-1){20}}  
  \put(136,122){$\mathbf{n=1}$}
  \put(-25,122){$1\otimes\bar{2}$} 
  \put(32,122){$2\otimes\bar{1}$}        
  \put(-12,118){\vector(1,-1){20}}
  \put(-18,118){\vector(-1,-1){20}}  
  \put(43,118){\vector(-1,-1){20}}
  \put(45,118){\vector(1,-1){20}}
  \put(136,87){$\mathbf{n=0}$}
  \put(186,87){$\mathbf{n'=2}$}
  \put(-55,87){$0\otimes\bar{2}$}       
  \put(3,87){$1\otimes\bar{1}$} 
  \put(58,87){$2\otimes\bar{0}$} 
  \put(-41,83){\vector(1,-1){20}}
  \put(13,83){\vector(-1,-1){20}}  
  \put(17,83){\vector(1,-1){20}}
  \put(69,83){\vector(-1,-1){20}}  
  \put(186,52){$\mathbf{n'=1}$}
  \put(-26,52){$0\otimes\bar{1}$} 
  \put(32,52){$1\otimes\bar{0}$} 
  \put(-10,49){\vector(1,-1){20}}
  \put(40,49){\vector(-1,-1){20}}
  \put(186,20){$\mathbf{n'=0}$}  
  \put(3,20){$0\otimes\bar{0}$}   
\end{picture} 
\end{center}
According to both methods non-quasi-primarity of the chiral level $k$ states induces non-quasi-primarity of the non-chiral level $\mathbf{k}$ state (here $\mathbf{n=1}$, $\mathbf{n'=1}$ respectively). Pursuant to QSC additionally states with lower level are affected (here the level $\mathbf{n=0}$ representative $1\otimes\bar{1}$). By contrast after KC the non-quasi-primarity is passed on states with higher level (here the $\mathbf{n'=2}$ state $0\otimes\bar{2}$+$1\otimes\bar{1}$+$2\otimes\bar{0}$). Two cases of the sketched scenario can be distiguinshed: 
\begin{itemize}
	\item Rank $r\geq3$: In KC non-quasi-primarity (NQP) of a chiral state with Jordan level $k$ ($\bar{k}$ respectively) encroaches upon all local states with $\mathbf{n'}\leq k$. In QSC the spread of NQP can be supressed by suitable choice of the considered representative unless the non-quasi-primary logarithmic partner is the field with highest Jordan rank. 
	\item Rank $r=2$: In KC the only possible non-quasi-primary is identical to the field with highest Jordan level, i.\,e.\ the NQP does not spread. In QSC every level zero representatives is affected by NQP. We deal with a situation that could be called converse quasi-primarity as the behaviour under $L_{0}$ and $\bar{L}_{0}$ equals the behaviour of a proper primary but for every representative 	\[\left|\mathbf{n=0}\right\rangle_{\alpha}=(1\otimes\bar{0}+0\otimes\bar{1})+\alpha(1\otimes\bar{1}-0\otimes\bar{1}),\; \alpha\in\mathbb{C}
\]
exists a positiv Virasoro mode such that
	\[\left(\bar{L}_n\right)^i\left(L_n\right)^j\left|\mathbf{n}=0\right\rangle_{\alpha}\neq0,\quad i,j\in\left\{0,1\right\},\quad i+j\neq0.
\]Actually, a promising possibility to cope with the spreading of NQP in the framework of QSC might be to extend $\mathcal{N}_{h\bar{h}}$. For an $r=2$ example we refer to \cite{GabR07}.
\end{itemize}

Exploring the first case might bring light to a problem we have already sketched in section \ref{loccondiss_sec}. It seems desirable to understand how the uncertainty can be mastered which enters computable data in the QSC formulation through the arbitrary but non-equivalent choice of the regarded representative. 

Furthermore the broken QSC--KC equivalence provides the opportunity to treat them as competing models. On the basis of an explicitly known realisation exhibiting the claimed properties it could become possible to determine wether the predicted spread of NQP is reasonable. And finally the comparison of the explicit realisation with the predictions of QSC and KC could resolve the question if one of the proposed methods can be adapted to the NQP case.  

\vspace{1cm} 
\textbf{Acknowledgements:} We would like to thank Matthias Gaberdiel, Kirsten Vogeler and Hendrik Adorf for valueable discussions. The work of MF is partially supported by the European Union network HPRN-CT-2002--00325 (EUCLID). 



\newpage
\begin{thebibliography}{10}
\bibliographystyle{}
\newcommand{\enquote}[1]{``#1''}

\bibitem{Gur93}
V. Gurarie.
\newblock \enquote{Logarithmic Operators in Conformal Field Theory}.
\newblock \emph{Nucl. Phys.}, \textbf{B410} (1993) 535--549.
\newblock {\tt [hep-th/9303160]}

\bibitem{Flo00}
M.~A.~I. Flohr.
\newblock \enquote{Null Vectors in Logarithmic Conformal Field Theory}.
\newblock \emph{JHEP Proc. Sect.}, \textbf{PRHEP-tmr2000/004}
\newblock {\tt [hep-th/0009137]}

\bibitem{Flo01}
M.~A.~I. Flohr.
\newblock \enquote{Operator Product Expansion in Logarithmic Conformal Field
  Theory}.
\newblock \emph{Nucl. Phys.}, \textbf{B634} (2002) 511--545.
\newblock {\tt [hep-th/0107242]}

\bibitem{GheKa97}
A.~M. Ghezelbash and V. Karimipour
\newblock \enquote{Global Conformal Invariance in D Dimensions and Logarithmic Correlation Functions}.
\newblock \emph{Phys.Lett.}, \textbf{B402} (1997) 282--289.
\newblock {\tt [hep-th/9704082]}

\bibitem{RAK96}
M.~R. Rahimi Tabar, A. Aghamohammadi and M. Khorrami.
\newblock \enquote{The Logarithmic Conformal Field Theories}.
\newblock \emph{Mod. Phys. Lett.}, \textbf{A12} (1997) 1349--1353.
\newblock {\tt [hep-th/9610168]}

\bibitem{RA05}
J.Rasmussen.
\newblock \enquote{On logarithmic solutions to the conformal Ward identities}.
\newblock \emph{Nucl. Phys.}, \textbf{B730} (2005) 300--311.
\newblock {\tt [hep-th/0507177]}

\bibitem{RA06}
J.Rasmussen.
\newblock \enquote{Affine Jordan cells, logarithmic correlators and hamiltonian reduction}.
\newblock \emph{Nucl. Phys.}, \textbf{B736} (2006) 225--258.
\newblock {\tt [hep-th/0508179]}

\bibitem{FloK05}
M.~A.~I. Flohr and M. Krohn.
\newblock \enquote{A Note on Four-Point Functions in Logarithmic Conformal Field Theories}.
\newblock \emph{Fortsch. Phys.}, \textbf{53} (2005) 456--462.
\newblock {\tt [hep-th/0501144]}

\bibitem{FloK03}
M.~A.~I. Flohr and M. Krohn.
\newblock \enquote{Four-Point Functions in Logarithmic Conformal Field Theories}.
\newblock \emph{Nucl. Phys.}, \textbf{B743} (2006) 276--306.
\newblock {\tt [hep-th/0504211]}

\bibitem{MRS02}
S. Moghimi-Araghi, S. Rouhani and M. Saadat.
\newblock \enquote{Use of Nilpotent weights in Logarithmic Conformal Field Theories}.
\newblock \emph{Int. J. Mod. Phys.}, \textbf{A18} (2003) 4747--4770.
\newblock {\tt [hep-th/0201099]}

\bibitem{Mathieu07}
P. Mathieu and D. Ridout
\newblock \enquote{From Percolation to Logarithmic Conformal Field Theory}.
\newblock {\tt [arXiv:0708.0802]}

\bibitem{GabK96}
M.~R. Gaberdiel and H.~G. Kausch.
\newblock \enquote{Indecomposable Fusion Products}.
\newblock \emph{Nucl. Phys.}, \textbf{B477} (1996) 293--318.
\newblock {\tt [hep-th/9604026]}

\bibitem{Gab01}
M.~R. Gaberdiel.
\newblock \enquote{An Algebraic Approach to Logarithmic Conformal Field
  Theory}.
\newblock \emph{Int. J. Mod. Phys.}, \textbf{A18} (2003) 4593--4638.
\newblock {\tt [hep-th/0111260]}

\bibitem{RasPear07}
J. Rasmussen and P.~A. Pearce
\newblock \enquote{Fusion Algebras of Logarithmic Minimal Models}.
\newblock {\tt [arXiv:0707.3189]}

\bibitem{EF06}
H. Eberle and M.~A.~I. Flohr.
\newblock \enquote{Virasoro Representations and Fusion for General Augmented Minimal Models}.
\newblock \emph{J. Phys.}, \textbf{A39} (2006) 15245--15286.
\newblock {\tt [hep-th/0604097]}

\bibitem{FloKn07}
M.~A.~I. Flohr and H. Knuth.
\newblock \enquote{On Verlinde-Like Formulas in $c(p,1)$ Logarithmic Conformal Field Theories}.
\newblock {\tt [arXiv:0705.0545]}


\bibitem{Flo97}
M.~A.~I. Flohr.
\newblock \enquote{Singular Vectors in Logarithmic Conformal Field Theories}.
\newblock \emph{Nucl. Phys.}, \textbf{B514} (1998) 523--552.
\newblock {\tt [hep-th/9707090]}

\bibitem{EbF05}
H. Eberle and M.~A.~I. Flohr.
\newblock \enquote{Notes on Generalised Nullvectors in logarithmic CFT}.
\newblock \emph{Nucl. Phys.}, \textbf{B741} (2006) 441--466.
\newblock {\tt [hep-th/0512254]}

\bibitem{GabK99}
M.~R. Gaberdiel and H.~G. Kausch.
\newblock \enquote{A local Logarithmic Conformal Field Theory}.
\newblock \emph{Nucl. Phys.}, \textbf{B538} (1999) 631--658.
\newblock {\tt [hep-th/9807091]}

\bibitem{Kau00}
H.~G. Kausch.
\newblock \enquote{Symplectic Fermions}.
\newblock \emph{Nucl. Phys.}, \textbf{B583} (2000) 513--541.
\newblock {\tt [hep-th/0003029]}

\bibitem{JPR06}
M. Jeng, G. Piroux and P. Ruelle.
\newblock \enquote{Height variables in the Abelian sandpile model: scaling fields and correlations}.
\newblock \emph{J. Stat. Mech.}, \textbf{0610} (2006) P015.
\newblock {\tt [cond-mat/0609284]}

\bibitem{GabR06}
M.~R. Gaberdiel and I. Runkel.
\newblock \enquote{The logarithmic triplet theory with boundary}.
\newblock \emph{J. Phys.}, \textbf{A39} (2007) 14745--14780.
\newblock {\tt [hep-th/0608184]}

\bibitem{GabR07}
M.~R. Gaberdiel and I. Runkel.
\newblock \enquote{From boundary to bulk in logarithmic CFT}.
\newblock {\tt [arXiv:0707.0388]}

\bibitem{SaSch05}
V.~Schomerus and H.~Saleur.\newblock \enquote{The GL(1|1) WZW-Model: From Supergeometry to Logarithmic CFT}.\newblock \emph{Nucl.~Phys.}, \textbf{B734} (2002) 211--245.
\newblock {\tt [hep-th/0510032]}

\bibitem{GQuSch06}
G. G\"otz, T.~Quella and V.~Schomerus.\newblock \enquote{The WZW model on PSU(1,1|2)}.\newblock \emph{JHEP}, \textbf{0703} (2007).
\newblock {\tt [hep-th/0610070]}

\bibitem{QuSch07}
T.~Quella and V.~Schomerus.\newblock \enquote{Free fermion resolution of supergroup WZNW models}.\newblock \emph{JHEP}, \textbf{0709} (2007).
\newblock {\tt [arXiv:0706.0744]}

\bibitem{Ly07}
A.--L. Do
\newblock \enquote{Diploma thesis}. 

\bibitem{DiFr97}
P. Di Francesco, P. Mathieu, and D. S\'en\'echal.
\newblock \enquote{Conformal Field Theory}.
\newblock Springer, 1997.

\bibitem{GurFl97}
V.~Gurarie, M.~A.~I. Flohr and C. Nayak.
\newblock \enquote{The Haldane-Rezayi Quantum Hall State and Conformal Field Theory}.
\newblock \emph{Nucl. Phys.}, \textbf{B498} (1997) 513--538.
\newblock {\tt [cond-mat/9701212]}

\bibitem{GabGod98}
M.~R. Gaberdiel and P. Goddard.
\newblock \enquote{Axiomatic Conformal Field Theory}.
\newblock \emph{Commun. Math. Phys.}, \textbf{209} (2000) 549--594.
\newblock {\tt [hep-th/9810019]}

\bibitem{Flo03}
M.~A.~I. Flohr.
\newblock \enquote{Bits and Pieces in Logarithmic Conformal Field Theory}.
\newblock \emph{Int. J. Mod. Phys.}, \textbf{A18} (2003) 4497--4592.
\newblock {\tt [hep-th/0111228]}

\bibitem{GSW87}
M.~B. Green, J.~H. Schwarz and E. Witten.
\newblock \enquote{Superstring Theory Vol 1: An Introduction}.
\newblock Cambridge University Press, 1987.

\bibitem{FGK07}
M. Flohr, C. Grabow and M. K\"ohn.
\newblock \enquote{Fermionic expressions for the characters of $c_{p,1}$ 
logarithmic conformal field theories}.
\newblock \emph{Nucl. Phys.}, \textbf{B768} (2007) 263--276.
\newblock {\tt [hep-th/0611241]}

\bibitem{CF06}
N. Carqueville and M. Flohr.
\newblock \enquote{Nonmeromorphic operator product expansion and 
$C_2$-cofiniteness for a family of ${\cal W}$-algebras}.
\newblock \emph{J. Phys. A: Math. Gen.}, \textbf{39} (2006) 951--966.
\newblock {\tt [math-ph/0508015]}

\end{thebibliography}
\end{document}